\documentclass{vgtc}                          

\graphicspath{{figures/}{pictures/}{images/}{./}} 

\usepackage{times}                     

\usepackage{tabu}                      
\usepackage{booktabs}                  
\usepackage{lipsum}                    
\usepackage{mwe}                       

\usepackage[T1]{fontenc}
\usepackage{csquotes}

\usepackage{mathptmx}                  

\usepackage{svg} 

\onlineid{1517}

\vgtccategory{Research}

\vgtcinsertpkg

\title{Experience Level Influences User's Criteria for Avatar Animation Realism}

\author{Yudong Huang\thanks{e-mail: yhuang143@hawk.illinoistech.edu}\\
\scriptsize Illinois Institute of Technology
\and Avneet Singh\thanks{e-mail: asingh180@hawk.illinoistech.edu}\\
\scriptsize Illinois Institute of Technology
\and Mark Roman Miller\thanks{e-mail: mmiller30@illinoistech.edu}\\
\scriptsize Illinois Institute of Technology}

\teaser{
  \centering
  \includegraphics[width=0.75\linewidth]{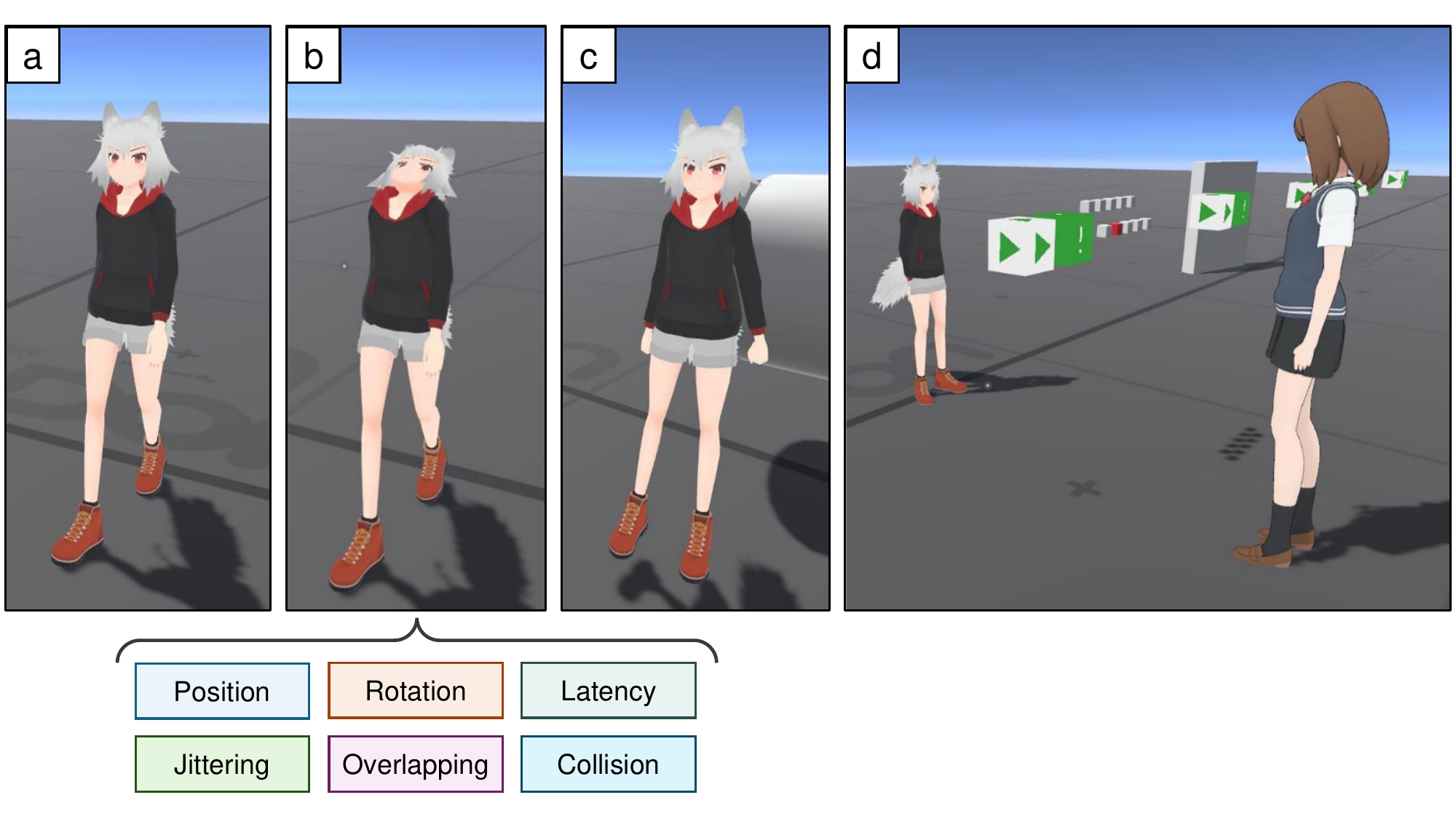}
  \caption{Scoring animation realism in VRChat. (a) Source animation version. (b) Modified animation version with six motion features (Jittering shown in the figure). (c) Reenacted animation. (d) Users play the animation and score its realism via buttons.}
  \label{fig:teaser}
}

\abstract{
The sense of realism in avatar animation is a widely pursued goal in social VR applications. A common approach to enhancing realism is improving the match between avatar motion and real-world human movement. However, experience with existing VR platforms may reshape users' expectations, suggesting that matching reality is not the only path to enhancing the sense of realism. This study examines how different levels of experience with a social VR platform influence users' criteria for evaluating the realism of avatar animation. Participants were shown a set of animations varying in the degree they reflected real-world motion and motion seen on the social VR platform VRChat. Results showed that users with no VRChat experience found animations recorded on VRChat unnatural and unrealistic, but experienced users in fact rated these animations as more likely to come from a real person than the motion-capture animations. Additionally, highly experienced users recognized the intent to imitate VRChat’s style and noted the differences from genuine in-platform animations. All these results suggest users' expectations of and criteria for realistic animation were shaped by their experience level. The findings support the idea that realism in avatar animation does not solely depend on mimicking real-world movement. Experience with VR platforms can shape how users expect, perceive, and evaluate animation realism. This insight can inform the design of more immersive VR environments and virtual humans in the future.

} 


\begin{document}


\firstsection{Introduction}

\maketitle

Social virtual reality (VR) is a form of social interaction within virtual reality where users engage with each other as avatars in immersive environments, emphasizing the integration of bodily movement, spatial presence, and emotional communication.

Sense of realism is a common goal in all VR applications, but especially so within social VR. People aspire to achieve experiences through VR that "seem very much like they are not mediated"~\cite{lombard1997heart}. In the context of Social VR, the realism of avatar animation is a key factor. It forms the foundation for effective social interaction in VR-embodied settings~\cite{schroeder2002social}, and is also an essential part of the sense of presence~\cite{capin1998realistic}. Existing research on avatar animation realism often focuses on making animations closer to real-world human activities—for example, enhancing visual quality~\cite{rogers2022realistic} or designing motion patterns that align with real-life logic~\cite{capin1998realistic}.

While it is often expected that Social VR platforms should provide perfect embodied animations to support user interaction, in reality, most avatar animations are far from flawless—yet this does not significantly diminish users’ sense of immersion and presence during social activities. For instance, VRChat allows users to track movement through headsets and controllers, leading to leg animations that are algorithmically generated and often appear stiff or even physically implausible. Nonetheless, users still willingly engage in social behaviors such as hugging and touching with these avatars. This suggests that the alignment between avatar animation and real-life movement is not the sole source of realism. Mel Slater and others have proposed the concept of plausible illusion, where the consistency between what users perceive and what they expect becomes a key component of realism~\cite{slater2009place}. In other words, if an avatar animation meets the user's expectations of what real users should look like on a given platform, it will be perceived as realistic. Crucially, real-world standards may not be the only basis for these expectations—experience within virtual environments can reshape users’ criteria for realism. Some studies in media theory have shown that even unreal content can, through psychological mechanisms, be internalized as part of a person's perception of social reality—even when users are fully aware that such content is fictional~\cite{shapiro1991making}.

Based on the above, we propose the following hypothesis: experience with a specific Social VR platform can reshape users’ criteria for what is considered “real.” Even if certain avatar animations appear unrealistic by real-world standards, users may come to accept—or even prefer—them due to their familiarity with the platform. This study focuses not on achieving higher fidelity environments, but on whether experience itself alters the evaluation criteria for realism. This perspective offers developers a new strategy: by intentionally designing platform-specific animation styles, they can actively shape users’ expectations of realism.

To test this hypothesis, the researchers designed a realism rating task within VRChat, where participants were asked to intuitively rate 64 avatar animations. These animations included 8 activity types and 8 animation types: high-fidelity motion capture animations representing real-world movement (Source animations), animations based on VR headset and controller tracking (Reenacted animations), and six types of Modified animations derived from the Source animations.

Participants were divided into two main categories: Non-users and Users, and further grouped into three experience levels—Low, Medium, and High. Results showed that Low-experience participants favored Source animations; Medium-experience participants rated Reenacted animations more favorably; and High-experience participants gave more similar scores, while demonstrating the ability to distinguish Modified animations mimicking VRChat style. These findings suggest that as experience increases, users develop platform-specific mental expectations, leading to more personalized and less reality-bound criteria for judging realism.

\section{Related Work}

\subsection{Social VR}
Social VR refers to a form of social interaction where users engage in real-time communication via virtual avatars in immersive three-dimensional environments using virtual reality technology. It is one of the application areas of VR technology. Platforms that implemented the concept of “virtual socialization” using computer technology began to emerge around the year 2000, such as Second Life launched by Linden Lab in 2003~\cite{SecondLife} and Roblox launched in 2005~\cite{Roblox}. These platforms were highly social and closely tied to avatars, but did not initially employ VR. It was not until the advent of consumer-grade VR devices in the 2010s that VR-based, multi-user embodied social platforms began to appear, such as AltspaceVR~\cite{AltspaceVR}, Meta Horizon Worlds~\cite{MetaHorizon}, and VRChat~\cite{VRChat}. Today, Social VR applications have developed sizable user communities.

Compared to traditional online socialization, Social VR has distinct affordances including embodiment, immersion, and richer interaction with the environment and space. Users can acquire digital surrogates (i.e., avatars)~\cite[p. 65]{ghaoui2005encyclopedia} in virtual environments, whose actions are directly controlled by the users. By using their own bodily movements to control avatars, users achieve a high degree of overlap between their real-world activities and their actions in virtual spaces. This leads to a stronger sense of immersion and greater freedom for creative expression~\cite{sykownik2021most}, offering new ways of interaction between people, and between people and environments~\cite{grabowski2024group, wei2022communication, hudson2019or}. These features provide a different experience from traditional social platforms. Beyond entertainment, academia and industry are also actively exploring application scenarios for Social VR, such as in education~\cite{hasenbein2022learning}, healthcare~\cite{li2023social}, and tourism~\cite{hudson2019or}.

\subsection{Realism in Social VR}



Realism in Social VR is a widely pursued goal. Surveys show that users clearly desire more immersive and realistic experiences in Social VR~\cite{sykownik2021most}. Moreover, highly immersive Social VR environments are also beneficial for sociological and psychological research, providing a stable and controllable setting that reflects participants’ genuine psychological states, making Social VR a valuable research platform~\cite{loomis1999immersive}.

Realism is an important yet ambiguous concept in the field of VR, with multiple interpretations and definitions. Sometimes, realism overlaps with other concepts such as presence and immersion. In most cases, it refers to the degree of trust users place in the content they experience in VR, and their tendency to respond positively to it. It indicates that users overlook the fictional nature of the VR setting and their own discretionary control, instead responding according to established schemas~\cite{bartlett1995remembering}, just as they would in real-world social contexts. The realism we investigate in this study is reflected in whether users feel that the animation they observe appears to originate from a real human. In this study, we focus on users’ experiences of avatar animation, and we refer to this perception as behavioral realism.


Behavioral realism is a component of the broader concept of realism, acting as one of the sources of such perception. Lombard proposed that the most fundamental aspect is the illusion of nonmediation~\cite{lombard1997heart}, meaning that users are unaware of the medium standing between themselves and the content. Slater distinguishes two aspects of presence: Place Illusion and Plausibility Illusion~\cite{slater2009place,slater2022separate}, and summarizes their combined result as the trigger for realistic behavior. Among them, Plausibility Illusion refers to the logical coherence of the scenario—for example, whether a character’s animation behavior aligns with common sense. In the work by Rogers et al.~\cite{rogers2022realistic}, behavioral realism was used as an experimental condition to study user responses. Our discussion centers on behavioral realism of avatars, defined as the extent to which a humanoid animation leads users to believe it is controlled by a real human.

While some studies define behavioral realism in terms of fidelity to real-world movement~\cite{rogers2022realistic,herrera2020effect,bailenson2006effect}, treating it as an objective variable, we emphasize that "observing an animation that closely resembles real human motion" and "believing the animation is from a real human" are two distinct phenomena. Users may observe animations that closely resemble real human motion but still judge them as not being produced by a real person. Conversely, users may also encounter animations that deviate from human kinematics yet still believe they are driven by a real human controller. For example, as noted in the Introduction, VRChat users do not feel uncomfortable when they see poor leg IK animations from their interaction partners. Therefore, we argue that closeness to real-life motion is not necessarily the only criterion for judging realism. We treat it as a dependent variable to be investigated through data collection. This marks a key difference between our study and other research on behavioral realism.

\subsection{Novelty and Experience Effects in VR}

During the use of VR, a unique phenomenon can be observed: as users become more familiar with a particular application or platform, they begin to adapt to the familiar animation styles. For example, some VRChat players may initially find avatar animations—tracked via headsets and controllers—rough or awkward, making it difficult to engage in social interaction. However, over time, as their usage increases, they become accustomed to this animation style and learn to interpret others’ body language. Eventually, socializing within this style no longer feels “out of place.” This study attributes such a phenomenon to the effects of experience and novelty on the judgment of realism. This phenomenon also supports the assertion discussed in Section 2.2 that observing human-like animations and perceiving them as controlled by real humans are not necessarily the same.

Heeter (1992)~\cite{heeter1992being} observed the influence of experience, stating that “having been there before can make it easier to believe you are there again.” Cox and  Yetter explored the effect of novelty on presence and immersion, showing that participants often experienced discomfort during their first VR sessions, but immersion and comfort increased as they became more familiar~\cite{cox2022so}. Jun also investigated the effect of novelty, with a focus on how participants responded to charitable donation requests after the experience~\cite{jun2023differential}. However, both studies emphasize the initial impact brought by first-time use. Grabowski et al. treated experience as a control variable to examine whether experienced users had better outcomes in VR-based meetings~\cite{grabowski2024group}. Other studies have also found that certain objective data from users changed over time with continued use~\cite{han2023people,miller2023large}.


The assumption that experience affects users’ psychological expectations, which in turn shapes their perception of realism, has theoretical support. Lombard ~\cite{lombard1997heart} mentioned that presence can be influenced by experience. Some literature in media and psychology has provided explanations. For instance, Shapiro et al. ~\cite{shapiro1991making} discussed how audiences might subconsciously internalize television content, explaining why people still accept narratives in soap operas despite knowing they are fictional. At the cognitive level, dynamic systems theory\cite[pp. 97--99]{newman2020theories}~\cite{fischer2006dynamic} proposes that human cognition is constantly evolving and being reshaped, which in turn influences how people perceive and interpret experiences. This study specifically focuses on the content of such dynamically evolving expectations and perceptual frameworks within the domain of humanoid animations on social VR platforms, namely, each user’s individual “criterion of realism.” It is important to clarify that although the term "criteria" is used collectively throughout this paper, this does not imply that all users share the same evaluative standards. Rather, this study adopts the premise that such criteria are internal components of users’ cognitive processes, rather than fixed, externally defined constructs.

\section{Method}

This section describes the programs and materials used in the experiment, aiming to understand users’ differences in their scoring criteria when assessing the realism of VRChat avatar animation. We begin by introducing the participants and describing the materials used, including devices, platforms, animation clips, and assigned tasks. We then outline the experimental design, procedure, and post-experiment questionnaire. Finally, we explain the data analysis methods, focusing on how scoring distributions varied by VRChat experience level.

\subsection{Participants} \label{ssec:participants}
\begin{figure}
    \centering
    \includegraphics[width=\linewidth]{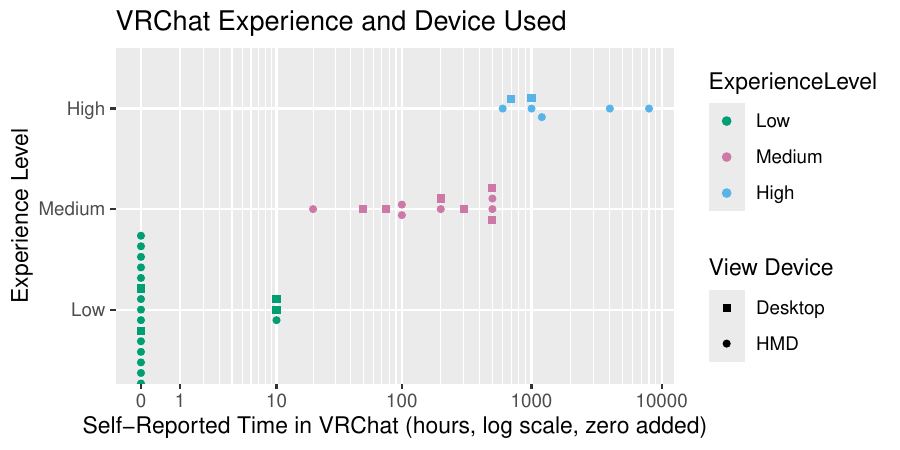}
    \caption{A beeswarm plot illustrating the distribution of self-reported time spent in VRChat. Each user is represented by a point with color and vertical position indicating the experience level, horizontal position indicating the self-reported time spent in VRChat, and shape representing the device used to access the virtual world.}
    \label{fig:experience}
\end{figure}

A total of 38 participants took part in the experiment, with one dropping out midway due to personal discomfort. In the end, 37 valid data samples were collected from a mix of 12 in-person and 25 remote participants. 
Among all valid participants, 9 were female and 28 were male. Participants’ ages ranged from 18 to 34 years, with a mean of 23.17 years and a standard deviation of 3.85. Fifteen participants had no prior experience with VRChat, while 22 participants had varying levels of experience. Based on their experience level, participants were further categorized into three participant groups: low experience (0–10 hours, 18 participants), medium experience (10–500 hours, 12 participants), and high experience (more than 500 hours, 7 participants). The distribution of time spent in VRChat, experience level assignment, and viewing device (described in Subsection \ref{sssec:apparatus}) is given in Figure \ref{fig:experience}.

Recruitment and experimental procedures were approved by the university's Institutional Review Board (IRB). Participants who agreed to the compensation policy received \$15 after completing the experiment. A within-subject design was used in this study, and all participants underwent the same experimental conditions.

\subsection{Materials}

\subsubsection{Apparatus and Environments}

\label{sssec:apparatus}

VRChat was used as the experimental platform. VRChat is a popular social VR platform that runs on several VR devices and is also compatible with PC. As a free platform, it has a large and active player community, with an average monthly concurrent player count consistently exceeding 20,000 over the past two years \cite{steamcharts}. Moreover, the platform allows users to create and upload their own avatars and virtual environments (referred to as worlds) in VRChat. These features make VRChat a suitable platform for conducting experiments and collecting data related to social VR.

Participants took part in the experiment either in person or remotely.
In-person participants used a Meta Quest 3 headset and controllers, with the headset weighing 515 grams. The device supports refresh rates of 90Hz/120Hz, a resolution of 2064 × 2208 per eye, and a field of view of 110° horizontally and 96° vertically. The headset was connected to the network via Wi-Fi during the experiment. The in-person setup also included a PC that ran VRChat and was connected to the same network as the headset via a wired connection. Running VRChat on the PC and streaming the content to the headset enabled richer features, higher performance, and easier export of experimental results. In-person participants stood at the center of a 3.6 m × 6.7 m room during the experiment.

Remote participants selected their own suitable environments and used their own familiar devices, including PCs, VR headsets and controllers, and full-body motion capture equipment to participate. The experimenter provided guidance and assistance via online voice communication.

The testing environment was developed using VRChat Creator Companion~\cite{VRChat} and Unity version 2022.3~\cite{Unity}. All experimental tasks were integrated into a single VRChat world and uploaded to VRChat. During the experiment, participants interacted with objects in the world to perform scoring, and the scoring data was automatically output to VRChat's logs through a custom script we developed in advance.

\subsubsection{Animations}

This study aimed to investigate how experience influences participants’ criteria for judging realism. The experiment compared scores across different animation types and experience levels, based on participants’ VRChat usage (see section \ref{ssec:participants}). Animations were grouped into three types: Source, Modified, and Reenacted, described below.

\textbf{Source animations} refer to high-fidelity motion capture animations obtained from sources outside VRChat. These animations closely match real-world human movements. In the analysis of the experimental results, this type of animation serves as a baseline score that reflects participants’ perception of regular human motion in VRChat. Since these animations serve as a reference for scoring, all other animation types share the same motion content, referred to as activities. Therefore, this type is named \textit{Source} animation. The resources used for Source animations came from the Unity Asset Store package MOBILITY PRO: Mocap Animation Pack\cite{MobilityPro}. According to its provider, Mocap Online, the animations were finely tuned based on motion capture data to ensure natural and smooth presentation in 3D environments.

\textbf{Reenacted animations} are avatar animations recorded directly in VRChat using a VR headset and controllers. In this study, researchers manually performed the motions from the Source animations to create these clips.  This method differs from PC-based scripted animations and full-body motion capture. Instead, it captures head and hand movement via the headset and controllers, while the rest of the body is estimated by VRChat’s algorithms.Though this form of animation may appear less realistic to new users, it is typical in VRChat and familiar to experienced users. This contrast supports the study’s goal of comparing perceptions across different experience levels. Animations were recorded using ShaderMotion, an open-source VRChat recording tool.

\begin{figure*}
    \centering
    \includegraphics[width=0.85\linewidth]{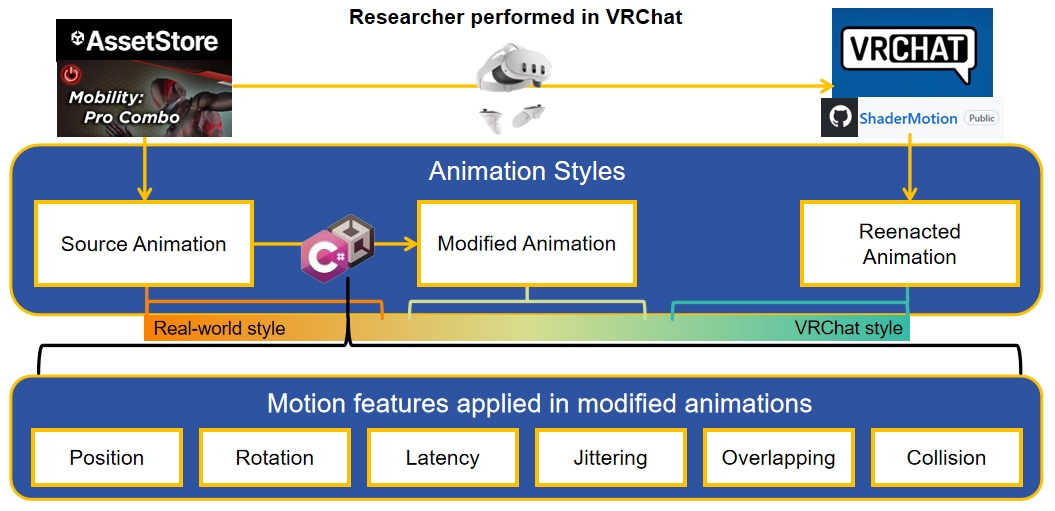}
    \caption{the preparation process for all animation styles. The Reenacted animation was performed by the researcher in VRChat using an HMD and hand controllers. The Modified animation was derived from the Source animation by editing keyframes via Unity C\# scripts to incorporate specific motion features. The three animation styles differ in how much they resemble real-world motion versus VRChat motion.}
    \label{fig:animation_styles}
    
\end{figure*}

\textbf{Modified animations} are Source animations altered using motion features taken from regular VRChat animations, which have distinct styles unlike real-world motion. These features were replicated and applied to make Source animations resemble Reenacted ones. Table \ref{tab:motion_features} lists all the motion features applied to the Source animations, including their names, description, cause, and methods of implementation. Keyframe edits were done via C\# scripts; others were applied manually in the virtual world. The three animation types are collectively referred to as animation styles, as illustrated in Figure \ref{fig:animation_styles}.

\begin{table}[tb]
  \caption{Avatar Animation Motion Features and Their Application}
  \label{tab:motion_features}
  \scriptsize
  \centering
  \begin{tabu}{X[2.2c] X[3.0c] X[3.0c] X[3.6c]}
    \toprule
    Motion Feature & Description & Cause & Implementation \\
    \midrule
    Position & Abnormal vertical position & UFO function* or uncalibrated position & Adjust vertical position of the model \\
    \midrule
    Rotation & Unnatural tilt & User tilted or horizon adjusted$\dagger$ & Adjust rotation value of the model\\
    \midrule
    Latency & Frame drops & Network issues & Hold one certain frame for several following frames \\
    \midrule
    Jittering & Limb jitter or unnatural motion & Tracking loss or device adjustment & Randomly vary limb parameters over frames \\
    \midrule
    Overlapping & Overlaps with environment & User's tracking device moved into object & Place the model overlapping an object \\
    \midrule
    Collision & Blocked by environment & Avatar moved into collider & Use the Source animation with fixed root position \\
    \bottomrule
  \end{tabu}
  {\raggedright *The UFO feature is a functionality in some avatars in VRChat that allows users to adjust the avatar’s height to make it appear as if it is floating in the virtual space. $\dagger$Horizon adjustment is a feature provided by VRChat, allowing users to modify the angle of the horizon, which may cause the avatar to appear tilted in the virtual space. \par}
\end{table}

Since we intend to investigate effects that generalize across activities, each animation type included eight types of activities. The activity types were consistent across all animation types. All activities are shown in Table \ref{tab:activities}.

\begin{table}[tb]
  \caption{Descriptions of Animation Activities}
  \label{tab:activities}
  \scriptsize
  \centering
  \begin{tabu}{X[1.2c] X[6.8c]}
    \toprule
    Activity & Description \\
    \midrule
    Stand & Model stands still with no large movements \\
    \midrule
    Crouch & Model crouches in place with no large movements \\
    \midrule
    Jump & Model jumps up and lands back in place \\
    \midrule
    Talk & Model stands and performs casual talking motions \\
    \midrule
    Relax & Model stands and moves limbs in a relaxed manner \\
    \midrule
    WalkF & Model walks forward \\
    \midrule
    RunF & Model runs forward \\
    \midrule
    RunB & Model runs backward \\
    \bottomrule
  \end{tabu}
\end{table}

All animations were played using the same open-source 3D character model\cite{Chita2022}. This model was sourced from Booth, where the author permits users to freely use, edit, and distribute it.

\subsubsection{Task}

In the Animations section, we presented 3 animation styles. Among them, the Modified animation was further divided into 6 categories based on the applied motion features. Including the Source animation and Reenacted animation, there were 8 different types in total.Each type contained 8 activities, resulting in 64 unique animations (8 types × 8 activities). Each animation was set as an individual scoring task. Participants were instructed to view each animation one by one and score its sense of realism, in order to evaluate their scoring under different combinations of type and activity. The relevant results were subsequently analyzed.

To reduce bias from comparing similar animations, the 64 tasks were divided into 11 groups (10 groups with 6 tasks, 1 with 4). While the tasks were intended to be shown in random order, a coding error caused all participants to see them in the same sequence. This limitation was addressed by including a term in the models representing the animation's position in evaluation order. Fortunately, given that 64 animations were evaluated, the selection of Source and Reenacted styles were interleaved and ordering appeared to have a minimal effect on the study outcome. Participants could rewatch and revise scores within each group, but once a group was submitted, it could not be reopened. New groups only appeared after the previous one was completed.

\subsection{Design and Procedure}
Throughout the experiment, participants were tasked with scoring the realism of animations presented in the VRChat experimental world, using buttons for all interactions. The experimental world contained only the animations and interaction buttons, without any other distracting elements. As previously mentioned, participants were required to complete a total of 64 scoring tasks, each consisting of one animation and one scoring session, grouped into sets of six.

\begin{figure}[ht]
    \centering
    \includegraphics[width=\linewidth]{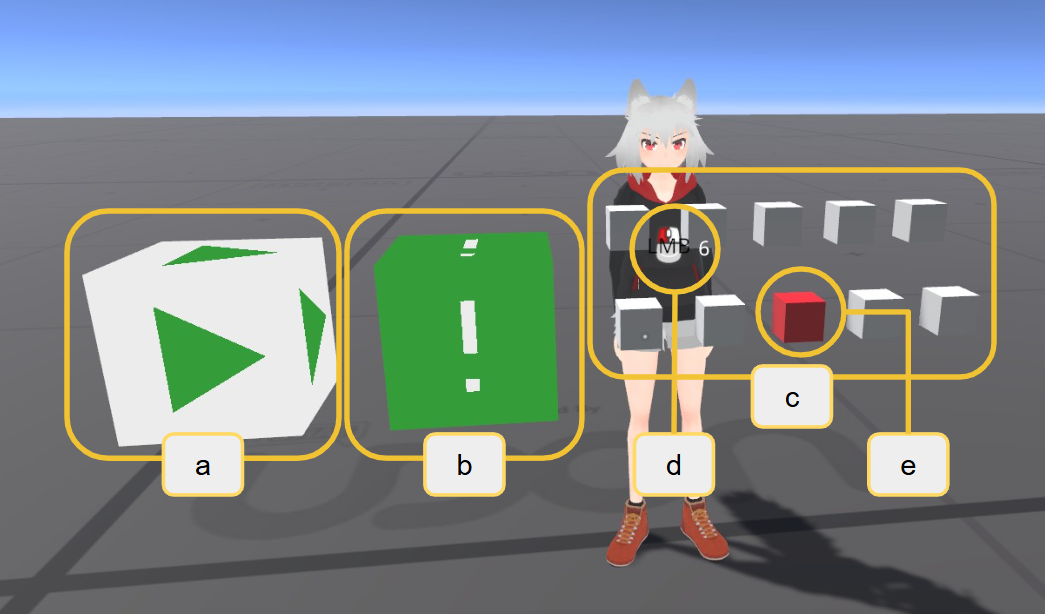}
    \caption{Button layout and interaction design in the experimental world: (a) Play Button. Click to display the character model and play the animation; (b) Scoring Button. Click to toggle the scoring panel on the right; click again to close it; (c) Scoring Panel. Each button corresponds to a score; (d) A tooltip appears at the location pointed to by the cursor or controller; (e) Previously selected scores are highlighted in red.}
    \label{buttons}
\end{figure}

Local participants navigated the experimental world using the default VRChat control scheme, moving with the left joystick and rotating their avatar orientation with the right joystick. They could also move and rotate physically in the real world to achieve the same effects. Remote participants used their preferred setup. Buttons allowed participants to view and rate each animation; previously selected scores were highlighted in red. Tasks in each set could be completed in any order, and new sets appeared after the previous one was finished. Figure \ref{buttons} shows the button interface.

Offline participants arrived at the lab, signed the consent form, and received in-person instructions and assistance from the experimenter while and after setting up the Quest 3 headset. Online participants signed the form via email and received a link to the VRChat world. The experimenter joined the same virtual world as online participants and explained the scoring tasks through VRChat’s voice system.

After learning the interaction and scoring procedures, all participants were instructed on how to evaluate the realism of virtual human animations. During the experiment, some Modified animations included effects that resembled software or hardware glitches, leading to occasional confusion. The experimenter clarified that all observed effects should be considered and encouraged participants to rely on their intuition, without highlighting any specific features.

After completing the final animation set, all animations and buttons disappeared. Offline participants were assisted in removing their headset; online participants exited the VR world on their own. All participants then completed a questionnaire on their scoring criteria, attention to six motion features, and realism judgments. Finally, the experimenter debriefed them and organized the collected data.

\subsection{Variables}
\label{ssec:variables}

Here we describe the variables that are part of the models.

\begin{itemize}
   \item \textbf{Participant:} Categorical variable unique per participant.
   \item \textbf{Usage:} Two categories: User and Non-user. This variable indicates whether the participant has prior experience with VRChat.
    \item \textbf{Experience:} Categorical variable representing finer-grained experience levels of Low, Medium, and High as illustrated in Figure~\ref{fig:experience}.
    \item \textbf{Animation Style:} Categorical variable including three types: Source, Reenacted, and Modified Animation.
    \item \textbf{Motion Feature:} A categorical variable describing modifications applied to Source animations. It includes six features, each representing a distinct animation feature found in VRChat that deviates from real-world motion. Together with the Source and Reenacted animation styles, this forms a total of eight animation types, as shown in Table~\ref{tab:motion_features}.
    \item \textbf{Activity:} Categorical variable including eight types, as shown in Table~\ref{tab:activities}.
\item \textbf{Realism:} The dependent variable in these studies was reported behavioral realism, reported on an ordinal scale from 1 to 10. The instructions given to the user were: "Your task is to rate all the animations in this world. Giving a high score means that you feel like you are watching the motion of a real human user." The phrasing, "you feel like you are watching" was adapted from the realism questions within Fraser et al. \cite{fraser2022expressiveness} which in turn was adapted from the Temple Presence Inventory~\cite{lombard2009measuring}. 
\end{itemize}




\subsection{Model}

In this work, we fit a linear mixed-effect model using the R package `lmerTest' version 3.1-3 \cite{lmerTest} wrapping the `lme4' package version 1.1-36 \cite{lme4}. This model was selected because there are multiple ratings by each participant and on each activity. There are likely correlations among ratings given by the same person, and there are potential correlations among ratings given to the same activity. These correlations violate the assumption of independence among observations that is necessary for standard linear models to have valid statistical inference. To address these concerns, we model both correlations with random effects.

In the process of producing this work, there were three other threats to the validity of these results: the linear analysis of Likert data that can be considered ordinal\cite{Norman2010}, participant location as a confounding variable, and view device as a confounding variable. We address these threats in the supplemental material, showing that each auxiliary analysis reaches the same conclusions about the hypotheses of this study.

\subsection{Hypotheses} \label{ssec:hypotheses}

This experiment proposes three progressively layered hypotheses. First, we test for an interaction effect to examine whether user experience has a non-substitutable influence on realism judgment. Second, we focus on experienced VRChat users to investigate whether they exhibit a consistent preference for certain animation styles (between Source animations and Reenacted animations). Finally, we examine Modified animations to determine whether specific motion features consistently receive higher realism ratings from the experienced VRChat user group.

\textbf{H1: Participants with less VRChat experience will rate animations with different \textit{criteria} than participants with more VRChat experience.} Note that the hypothesis is not that experienced participants will give higher (or lower) ratings across all animations, but that there will  be certain motions rated relatively higher by one group that are rated relatively lower by another group, and vice versa. This can be evidenced by an interaction effect between the animation style (i.e., Source vs. Reenacted animations) and experience.

\textbf{H2: Participants with more VRChat experience will rate Reenacted animations to be more realistic than Source animations.} H1 only indicates that there is an interaction effect; H2 in contrast challenges the idea that motions representing real-world motion are always the most realistic. H2 anticipates that, within the specific context of social VR, certain animation styles may evoke stronger sense of realism among experienced users. 

\textbf{H3: Modifying Source animations with motion features simulating VRChat errors will increase the realism of the animations for experienced VRChat users.} If H3 is supported, it would offer VR application and platform designers at least one strategy for making animations provide more realism to VRChat users.

\section{Results}

In this study, we present three models examining the effects of user experience in VRChat on the perception of avatar realism. These models vary slightly in model terms and data used. The first two models, sections \ref{ssec:usage} and \ref{ssec:experience}, provide two ways in which we address H1 and H2. The models vary in how previous experience is operationalized. The model within Section \ref{ssec:usage} uses Usage but in \ref{ssec:experience} the model uses Experience (see \ref{ssec:variables} for definitions). These models only examine ratings of Reenacted and Source animations, and do not include Modified animations. The third model, Section \ref{ssec:modified}, aims to answer H3, uses all rating data, and uses the Experience variable.  The models follow the hypotheses described in subsection \ref{ssec:hypotheses}, and each model's variables and data are described below.

\subsection{Usage on Realism Criteria}

\label{ssec:usage}

\begin{figure}[tb]
    \centering
    \includegraphics[width=\linewidth]{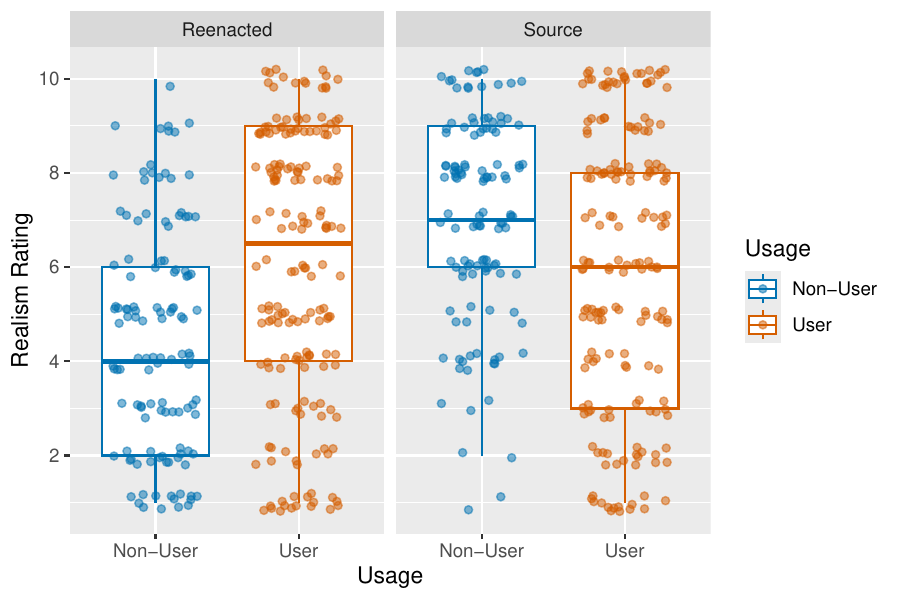}
    \caption{A plot of participant's ratings of realism of an animation broken down by the motion within the animation (Source or Reenacted) and the participant's usage of VRChat (User vs. Non-User). Each dot represents one rating, jittered for visibility. Boxplots show median and upper and lower quartiles.}
    \label{fig:model1_raw}
\end{figure}

In this first analysis, we focus solely on Source and Reenacted styles, and divide participants into the two pre-defined categories as established by the "Usage" variable. The model estimated the participant’s rating of each agent’s animation with fixed effects of style (Reenacted or Source), VRChat usage, a linear term accounting for evaluation order, and an interaction term between style and VRChat usage. The random effects were a random intercept for each participant and a random intercept per animation type. The rating points along with median and inter-quartile range are shown in Figure \ref{fig:model1_raw}.

In the prototypical case (i.e., with random effects set to zero) the effect of style was estimated at a -0.829 point decrease in realism on the ten-point scale for the VR case relative to the average case ($t(542.988)=-7.597, p < 0.001$). The model estimated the effect of usage to be 0.048 point increase in realism on the ten-point scale, though it was not significant itself ($t(34)=0.212, p = 0.833$). The model also estimated the effect of evaluating on HMD to be a -0.234 point decrease in realsim on the ten-point scale, though it also was not significant itself ($t(34)=-0.993, p = 0.328$). The effect of evaluation order was not significant, estimated to be 0.002 point increase in realism per evaluation performed ($t(15.436)=0.342, p = 0.737$).

However, there were two interaction effects found. The first is most important to this work as it confirms hypothesis \textbf{H1}: there was an interaction effect between VRChat usage and style such that Non-users viewing original motion and users viewing Reenacted motion were predicted to have a 0.800 point increase in realism relative to the average case ($t(542.233)=7.605, p < 0.001$). Second, there was an interaction effect between view device and style such that users viewing in an HMD and non-users viewing on desktop were predicted to have a 0.487 point increase in realism relative to the average case ($t(542.233)=4.413, p < 0.001$).

The standard deviation of the random effect of participant was 1.118, and the standard deviation of the random effect of animation was 0.313.

We then performed post-hoc tests to understand the interaction effect between usage and motion style. As illustrated by Figure \ref{fig:model1_raw}, there was not a significant difference in the ratings of Reenacted vs. Source motion styles by VRChat users ($z=0.045, p=0.944$), failing to confirm hypothesis \textbf{H2} in this initial analysis. However, in the following section, we do find evidence confirming \textbf{H2}.

\subsection{Experience Level on Realism Criteria}

\label{ssec:experience}
 
\begin{figure}
    \centering
    \includegraphics[width=\linewidth]{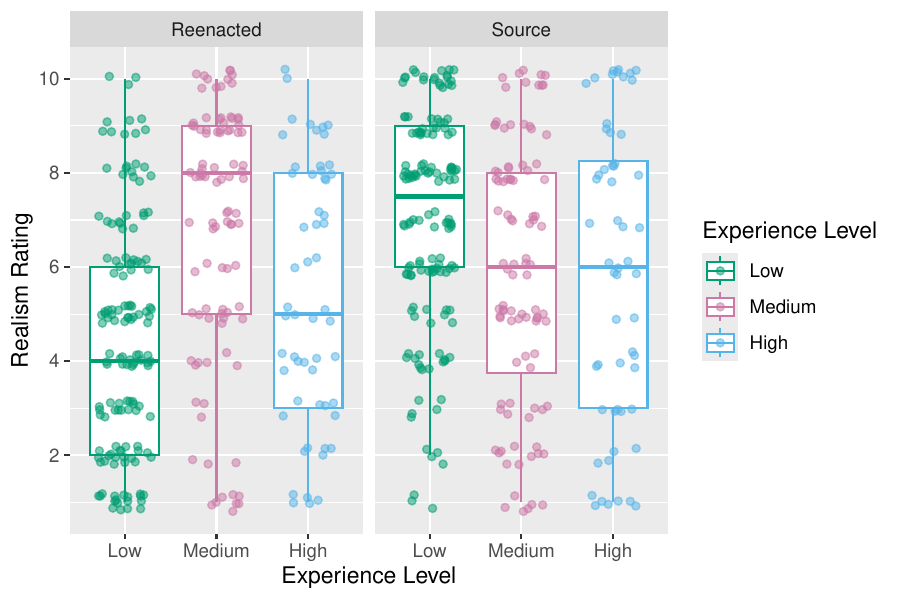}
    \caption{A plot of participant's ratings of realism of an animation broken down by the motion within the animation (Source or Reenacted) and the participant's degree of experience with VRChat (Low vs. Medium vs. High). Each dot represents one rating, jittered for visibility. Boxplots show median and upper and lower quartiles.}
    \label{fig:model2_raw}
\end{figure}

The second model distinguishes users not simply between Users and Non-users but into three levels of experience, low, medium, and high, as described in subsection \ref{ssec:participants}. This second analysis also uses only Source and Reenacted styles like the first model. The model estimated the participant’s rating of each agent’s animation with fixed effects of style (Reenacted or Source), VRChat experience (low, medium, and high), a linear term accounting for evaluation order, and an interaction term between style and VRChat experience. The random effects were a random intercept for each participant and a random intercept per animation type. The rating points along with median and inter-quartile range are shown in Figure \ref{fig:model2_raw}.

With the experience variable being categorical, the Medium and High variables were dummy-coded and the Low experience level was used as the baseline value. In this case, the effect of style was estimated at a -1.478 point decrease in rated realism on the ten-point scale for the VR case relative to the average case ($t(541.579)=-9.848, p < 0.001$). The model estimated the effect of Medium experience relative to baseline to be a 0.442 point increase in rated realism, though it was not significant itself ($t(33)=0.907, p = 0.371$). The model also estimated the effect of High experience relative to baseline to be a -0.133 point decrease in rated realism, which was also not significant ($t(33)=-0.237, p = 0.814$). The effect of evaluation order was not significant, estimated to be a 0.002 point increase in rated realism per evaluation performed ($t(15.825)=0.311, p = 0.76$).

Both interaction terms of style with both medium and high experience levels were significant. First, there was an interaction effect between Medium (vs. Low) experience level and style such that Medium experience users viewing Reenacted motion and Low experience users viewing Source motion were predicted to have a 1.904 point increase in rated realism relative to the average case ($t(541.184)=8.428, p < 0.001$). Second, there was an interaction effect between High (vs. Low) experience level and style such that High experience users viewing Reenacted motion and Low experience users viewing Source motion were predicted to have a 1.068 point increase in rated realism relative to the average case ($t(541.184)=4.092, p < 0.001$). Both of these interactions are visible in Figure \ref{fig:model2_raw}.

Similar to the simpler model, there was also an interaction effect between view device and style. Users that viewed the VRChat-captured animations in an HMD and the mocap animations on Desktop were predicted to score the avatar with a 0.434 point increase in rated realism relative to the average case ($t(541.184)=4.075, p < 0.001$).

The standard deviation of the random effect of participant on realism rating was 1.12, and the standard deviation of the random effect of activity on realism rating was 0.318.

With interaction terms being significant, we again apply post-hoc tests with `emmeans’ to clarify these relationships. First, we find that among Low-experience participants, Reenacted style animations are rated significantly lower in realism than Source style by 2.957 on the ten-point scale ($z=-9.847, p < 0.001$). Within Medium-experience participants, Reenacted styles are rated significantly higher in realism than Source motion by 0.852 on the ten-point scale ($z=2.522, p = 0.012$). Finally, within High-experience participants, Reenacted style are rated lower in realism than Source motion by 0.821 on the ten-point scale ($z=-1.817, p = 0.07$).

While these tests can indicate whether each group rates Source animations different from Reenacted animations, that fact does not necessarily confirm the criteria vary according to group. This is done by a test of the interaction term, comparing the difference between Source and VRChat from group to group. These interaction contrasts were performed with emmeans as well.

The ratings of Low-experience participants of Reenacted style and Medium-experience participants on Source style were significantly lower in realism than Low-experience participants of Source style and Medium-experience participants on Reenacted style by an estimated value of 3.809 on the ten-point realism scale ($z=-8.428, p < 0.001$). The ratings of Low-experience participants of Reenacted style and High-experience participants on Source style were significantly lower in realism than Low-experience participants of Source style and High-experience participants on Reenacted style by an estimated value of 2.136 on the ten-point realism scale ($z=-4.092, p < 0.001$). The ratings of Medium-experience participants of Reenacted style and High-experience participants on Source style were significantly lower in realism than Medium-experience participants of Source style and High-experience participants on Reenacted style by an estimated value of 1.673 on the ten-point realism scale ($z=2.966, p = 0.003$). The directions and strength of these comparisons are visible in Figure \ref{fig:model2_raw}.

Using this second operationalization of experience, we were able to confirm both hypothesis \textbf{H1} with the interaction effects and hypothesis \textbf{H2} by focusing on the Medium-experience group in particular.

\subsection{Modified Motions}

\label{ssec:modified}

\begin{figure}
    \centering
    \includegraphics[width=\linewidth]{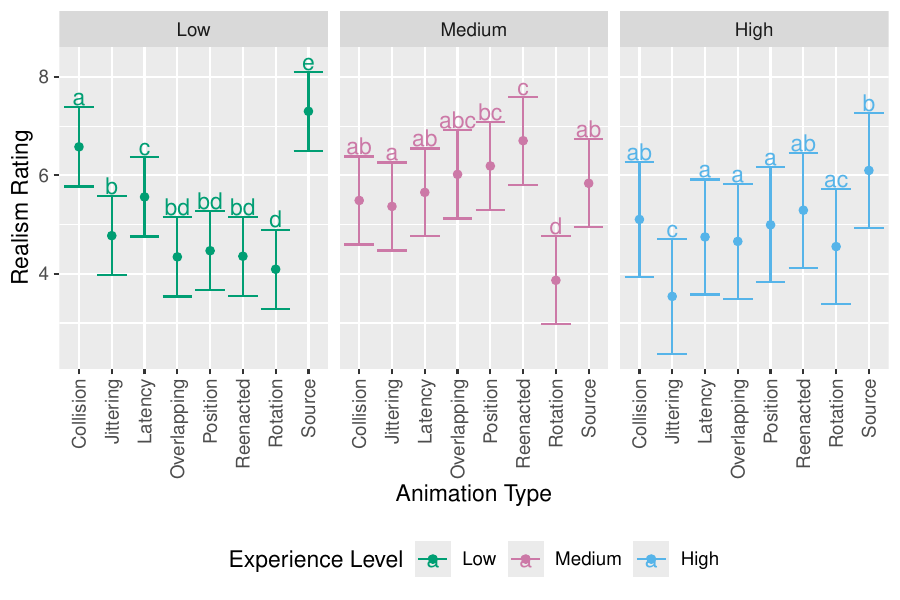}
    \caption{A plot expressing the point estimates of realism of an animation broken down by the motion within the animation (eight types) and the participant's degree of experience with VRChat (Low vs. Medium vs. High). Error bars represent 95\% confidence intervals. Within each facet, estimates that are significantly different at $\alpha<0.05$ do not share any letters labeling the error bar.}
    \label{fig:model3_type}
\end{figure}

The third model retained the division between the three levels of experience, but increased its scope to modified Source motions as well. The model estimated the participant’s rating of each agent’s animation with fixed effects of animation type (now one of 8 animation types), VRChat experience (Low, Medium, and High), a linear term accounting for evaluation order, and an interaction term between motion type and VRChat experience. The random effects were a random intercept for each participant and a random intercept per activity type. The fitted values are displayed in Figure \ref{fig:model3_type}. Given the number of comparisons, instead of each statistical test being listed off, significant comparisons are described using letters such that any bars within a plot's facet are significantly different if they do not share any letters. Significance is defined at the $\alpha < 0.05$ level, corrected for multiple comparisons with false-discovery rate.

Of interest in these comparisons are whether any Modified animation (i.e., an animation that is neither Source or Reenacted) can attain a greater motion realism rating than the Source animation. In considering both the Medium and High experience level panels in \ref{fig:model3_type}, no modified motion has a significantly greater realism estimate than Source. The closest is Position vs. Source in the Medium category, where the estimated realism of Position was indeed higher than Source by 0.352 on the realism scale, but not significantly so ($t(2295)=1.044$, uncorrected $p=0.2694$). Ultimately, we failed to confirm \textbf{H3}.

\section{Discussion}

This experiment examined the influence of VRChat experience on the judgment of realism of avatar animations. Our results indicated that experience does change the criteria users have for realistic motion (confirming \textbf{H1}). When dividing the population into two groups, Users and Non-users, we did not find that the Reenacted style was rated as more realistic than the Source style. However, when we worked with three levels of experience, we found that virtual motions with higher fidelity to real-world motions are rated as less realistic than motions more present in VRChat (confirming \textbf{H2}). Finally, we were not able to artificially modify Source animation to have higher ratings from the experienced participants (failing to confirm \textbf{H3}).

\subsection{VRChat Users have Different Realism Criteria Than Non-Users}

Our results indicate that participants with different experience levels adopted different scoring criteria when evaluating the realism of avatar animations. When dividing participants into Users and Non-users, we observed a significant interaction effect: non-users rated Source animation as more realistic and gave significantly lower scores to Reenacted animation. In contrast, user participants showed greater tolerance and acceptance toward animations in the VRChat style, suggesting that their judgment of "realism" relied more on platform familiarity than on alignment with real-world movement. It is worth noting that some participants exhibited a form of "overfitting" during their evaluations. For example, even when the latency presented in the Modified animations did not fully match the delay effects typically experienced in VRChat, some participants did not recognize the discrepancy and instead reported in the questionnaire that such latency was characteristic of VRChat itself. This indicates that although imperfect animations' experience shapes users' reasoning and criteria in the first place, the criteria derived from their experience eventually evolve into a judgment system that extends beyond the boundaries of the original experiential context.

According to in-experiment feedback and post-study questionnaires, non-user participants frequently emphasized the importance of smoothness and physical plausibility in the animations. For example, a non-user participant criticized Modified animations with the "Rotation" motion feature, commenting that "only Michael Jackson could move like that," referring to the pop artist Michael Jackson's nearly forty-five degree tilt in the music video for ``Smooth Criminal.'' User participants were more concerned with whether the animations were achievable within VRChat. They would instinctively attempt to replicate the actions shown in the animation. Several user participants mentioned that their main criterion was whether they or their friends had previously achieved that kind of effect in VRChat.

Overall, these results indicate that the realism of avatar animations is not a fixed standard but a cognitive construct that evolves with the accumulation of user experience and shifting expectations within virtual platforms.

\subsection{High-Fidelity Animations Can Seem Less Realistic to Medium-Experience Participants}

Our study found that participants with medium experience rated Source animation as less realistic than Reenacted animation. This suggests that the sense of realism for Medium-experience users is shaped by internalized cognitive expectations developed through platform use. For these users, frequent exposure to VRChat motion styles based on headsets and controllers has reshaped their standards for what counts as “realistic behavior.” As a result, animations that closely follow real-world physical principles may appear jarring or unnatural within the virtual context they are familiar with. For example, some Medium-experience participants noted in the questionnaire that the running speed in the RunF animation of the Source animations was too fast—something rarely seen in VRChat.

This also helps explain why Medium-experience participants gave lower scores to Source animation compared to Low-experience users: in VRChat, not only are certain VRChat-style motions frequently seen, but some motions common in the real world are rare or even impossible. In contrast, Low-experience participants lack this knowledge and focus mainly on smoothness and resemblance to real-world movement, making them unaware to this discrepancy.



This effect bears some relationship to the uncanny valley effect \cite{mori2012uncanny}. The uncanny valley is the phenomenon that characters that are almost human in appearance become off-putting. There are several theories for this, but it is commonly believed the source of this is a mismatch between a highly realistic appearance but unnatural motion. It is plausible that the photorealism of the character sets expectations for realism within the context of the character that the behavior does not match, just as our Medium-experience participants had expectations for behavior that were not met. There is a difference between these two cases because the uncanny valley actually produces a negative emotion (disgust, fear) whereas we did not find evidence of that in our study. Whether this similarity is true or not is beyond the scope of this study, but it is worth noting that experience does not simply mean experience of a platform, but could extend to other patterns in stimuli such as characters.

\subsection{High-Experience Participants Had Finer Realism Criteria}

High-experience participants gave similar realism ratings to both animation styles, showing a more neutral and complex evaluation. This likely reflects their broader exposure to various animation types in VRChat, including high-fidelity motion capture.

Many High-experience participants are dedicated VRChat users, with access to high-end motion capture equipment and active user communities. These factors increase their chances of being exposed to high-fidelity motion capture animations, shaping more flexible criteria for realism and reducing bias toward any specific animation style.

Another supporting observation is that High-experience participants could distinguish between animations that are native to the VRChat platform and those that imitate Source animations. During post-study interviews, several High-experience participants made comments such as:

\begin{displayquote}

“Frame drops in VRChat don’t jump to a still frame after freezing; instead, they move very rapidly to the post-lag position, and that movement still carries inertia.”

“Tracking loss usually happens when the hands leave the camera’s field of view. When the hands are at the sides or behind the body, tracking loss is actually less likely.”

\end{displayquote}

This phenomenon—where realism becomes harder to attain as experience increases—also echoes the idea discussed by Lombard et al.~\cite{lombard1997heart}, which suggests that over-familiarity can undermine presence: “An engineer cannot help but notice flaws in a virtual environment or the image in a high definition television system because she/he either knows or wants to know what is responsible for the flaw; this knowledge reminds her/him that the experience is mediated.” However, it is important to clarify that this does not imply that increased experience inevitably leads to the loss of realism. Rather, it indicates that the user’s standard for realism becomes highly specific—demonstrating the profound influence of experience on the criteria for evaluating realism.

Notably, some animations that are theoretically plausible but not commonly seen in VRChat received higher scores from High-experience participants than from both Low- and Medium-experience groups. This caused certain motion features to show a “rebound” in scores across experience levels. The Rotation motion feature in Figure \ref{fig:model3_type} illustrates this pattern. One High-experience participant remarked, “People don’t usually do this on purpose, but you can actually adjust the horizon in VRChat to make it happen.”

This again confirms that High-experience participants internalize their understanding of the platform as part of their realism evaluation—just as Low-experience participants rely on real-world physical laws as their standard. This reflects the cognitive process of experience being internalized into realism judging criteria.

\subsection{Implications for Theory }

This study offers a new perspective on the concept of “realism” in virtual environments. Building on the common idea that realism stems from user expectations, we propose that users’ criteria for realism can evolve with experience. The expected “reality” is not a static reflection of the real world, but a cognitive construct shaped through repeated exposure.

This complements the traditional view that realism increases with similarity to real-world motion. Our results show that judgments of realism are highly context-dependent and influenced by platform experience. While Low-experience users rely on real-world standards, experienced users develop evaluation criteria that align more closely with the unique characteristics of the platform.

As evaluation standards shift, so do the perceived realism and presence. This suggests that Social VR platforms need not replicate reality to be immersive—they can define their own version of reality. Presence, in this view, is not universal but shaped by technical expressions and community culture.

In sum, this study highlights how experience reshapes users’ standards for realism, and calls for a broader understanding of both “realism” and “presence.” In Social VR, realism is co-constructed through the dynamic interaction between user expectations and platform conventions.

\subsection{Limitation and future work}





Several limitations should be considered when interpreting the results of this study.

First, participants were recruited on a voluntary basis, which may introduce selection bias. For example, individuals with higher social motivation or greater interest in VR social interaction may have been more likely to participate, potentially affecting the generalizability of the findings. The recruitment also meant that there was self-selection by experience level (it was not randomly assigned) so these findings could not distinguish between the effect of experience and other causes that influence adoption or retention.

Second, the animation clips used in the experiment were relatively short and lacked contextual information, which may have made it difficult for participants to infer behavioral intent and thus impacted their judgments of realism.

Third, the study did not fully control for the influence of experience on other platforms. Some participants categorized as having low experience may have previously engaged with other platforms, which could have shaped their expectations and scoring. For instance, some participants familiar with MMO games found the collision feature in the Modified animations to be natural.

Finally, this study focused on six common motion features in VRChat to keep the task manageable for participants. While practical, this selection may not fully represent all animation styles generated through headset and controller-based motion capture Modified animations, created via C\# scripts, are not perfect replications of authentic VRChat movements. Therefore, H3 may still be supported in future studies with more precise reconstructions.



Future research could explore how users assess realism in longer, narrative-driven animations and consider demographic factors like age. It may also use more technically representative VRChat animations, such as applying inverse kinematics to Source animations. Extending the study to other mainstream platforms could test the generalizability of these findings.

\section{Conclusion}
Our hypothesis is that the perception of avatar realism differs between VRChat users and non-users. The level of experience within virtual environments, such as VRChat, can influence an individual's ability to perceive an avatar as representing a real person.


Our interview findings indicate that VRChat users that had high experience (i.e, those with 500 hours or more) were able to distinguish between animation styles that we had modified and those captured by VRChat. This suggests that user experience plays a critical role in shaping realism perception. Although the failure to confirm H3 means that the experiment could not identify a specific motion feature capable of directly transforming Source animations into versions that experienced VRChat users perceive as realistic, the confirmation of H1 and H2 still provides valuable guidance for VR designers and developers. Specifically, the implementation of animations in VR does not need to adhere strictly to "realism" in the traditional sense. Platforms have the capacity to shape user preferences through consistent exposure and stylistic reinforcement, leading users to accept—and even favor—certain animation styles.

This insight encourages the development of stylized animations with strong platform-specific cultural identities, allowing each platform to cultivate its own unique language of immersion. It also grants developers and content creators greater creative autonomy in defining the aesthetic and experiential character of their virtual environments.


\acknowledgments{
The authors thank the Social Spatial Interaction Laboratory at Illinois Institute of Technology for their support, Kaiang Wen for assistance with figures, and the VRChat players worldwide for participating in the study.
}

\bibliographystyle{abbrv-doi}

\bibliography{template}
\end{document}